\documentclass[useAMS]{mn2e}
\usepackage{amssymb,amsmath}
\usepackage{graphicx}

\title[3C390.3: disk-like BLRs near central BH]
{3C390.3: More stable Evidence for Origination of Double-Peaked Broad 
Balmer Lines from Accretion Disk Near Central Black Hole}
\author[Zhang X.-G.]
       {Xue-Guang Zhang$^{1,2}$\thanks{xgzhang@pmo.ac.cn}\\
       $^1$Purple Mountain Observatory, Chinese Academy of Sciences,
             2 Beijing XiLu, Nanjing, Jiangsu, 210008, P. R. China\\
       $^2$Department of Physics and Astronomy, Texas A\&M University, 
             College Station, Texas, 77843-4242, U.S.A.}

\date{}
\def\LaTeX{L\kern-.36em\raise.3ex\hbox{a}\kern-.15em
    T\kern-.1667em\lower.7ex\hbox{E}\kern-.125emX}

\begin{document}
\pagerange{\pageref{firstpage}--\pageref{lastpage}} \pubyear{2010}
\maketitle
\label{firstpage}

\begin{abstract}
    In this manuscript, the structure of broad emission line regions (BLRs) 
of well-mapping double-peaked emitter (AGN with broad double-peaked 
low-ionization emission lines) 3C390.3 is studied. Besides the best 
fitted results for double-peaked broad optical balmer lines of 3C390.3 by 
theoretical disk model, we try to find another way to further confirm the 
origination of double-peaked line from accretion disk. Based on the long-period 
observed spectra in optical band around 1995 collected from AGN WATCH 
project, the theoretical disk parameters of disk-like BLRs supposed by 
elliptical accretion disk model (Eracleous et al. 1995) have been well 
determined. Through the theoretical disk-like BLRs, characters of observed 
light-curves of broad double-peaked H$\alpha$ of 3C390.3 can be well 
reproduced based on the reverberation mapping technique. Thus the accretion 
disk model is preferred as one better model for BLRs of 3C390.3. Furthermore, 
we can find that different disk parameters should lead to some different 
results about size of BLRs of 3C390.3 from the one measured through 
observational data, which indicates the measured disk parameters are 
significantly valid for 3C390.3. After that, the precession of theoretical 
elliptical disk-like BLRs being considered, we can find that the 
expected line profile in 2000 by theoretical model is consistent with 
the observed line profile by HST around 2000. Based on the results, we 
can further believe that the origination of broad double-peaked balmer 
emission lines of 3C390.3 are from accretion disk around central black hole.
\end{abstract}


\section{Introduction}

    The properties of objects with broad double-peaked low-ionization 
emission lines (hereafter, double-peaked emitters) have been studied 
for more than two decades, since the double-peaked emitters were found 
in nearby radio galaxies (Stauffer et al. 1983, Oke 1987, Perez et al. 
1988, Halpern 1990, Chen et al. 1989, Chen \& Halpern 1989). Some 
theoretical models have been proposed to explain the properties of 
double-peaked broad emission lines which can be used as one probe 
of the broad-line regions of active galactic nuclei (Eracleous et al. 
2009), such as binary black hole model (Begelman et al. 1980, 
Gaskell 1983, 1996, Boroson \& Lauer 2009, Lauer \& Boroson 2009, 
Zhang et al. 2007), double stream model (Zheng et al. 1990, 1991, 
Vellieux \& Zheng 1991), accretion disk model (Chen et al. 1989, 
Chen \& Halpern 1989, Eracleous et al. 1995, 1997, Bachev 1999, 
Hartnoll \& Blackman 2000, 2002, Karas et al. 2001, Gezari et al. 
2007, Flohic \& Eracleous 2008, Lewis et al. 2010, Tran 2010, 
Chornock et al. 2010, Gaskell 2010) etc.. Although which theoretical 
model is more preferred for double-peaked emitters is still an open 
question, the accretion disk model which indicates double-peaked 
broad emission lines are from central accretion disk is so far the 
more widely accepted model applied to explain the properties of 
double-peaked broad emission lines. 

    Due to the un-obscured emitting region on the receding jet, 
the double-stream model can be ruled out for double-peaked emitter 3C390.3 
(Livio \& Xu 1997). Through the long-period observational results, 
binary black hole model has also been ruled out due to unreasonable 
large central BH masses for some double-peaked emitters, such as Arp102B, 
3C390.3 (Eracleous et al. 1997). Furthermore, based on the long-period 
variations of double-peaked broad emission lines, accretion disk model is 
successfully applied to reproduce the characters of varying observed 
line profiles of double-peaked emitters, such as the individual 
double-peaked emitter NGC1097 (Storchi-Bergmann et al. 1995, 1997, 2003), 
and one sample of double-peaked emitters (Lewis et al. 2010, Gezari et al. 
2007). However, we should note that although accretion disk model is 
so far the widely accepted theoretical model for double-peaked emitters, 
the other theoretical models can be expected to be preferred for some 
special double-peaked emitters, such as the x-shaped radio objects 
which could be candidates for objects with binary black holes in 
central regions (Merritt \& Ekers 2002, Zhang et al. 2007). 
Thus, only based on the double-peaked appearance of broad emission 
lines, there is no confirmed way to affirm which theoretical model 
is more preferred for double-peaked emitter.

    Among the sample of double-peaked emitters (Eracleous \& Halpern 
1994, 2003, Eracleous et al. 1995, Strateva et al. 2003), 3C390.3 is 
one well studied double-peaked emitter, also one well-studied mapping 
AGN (Dietrich et al. 1998, O'Brien et al. 1998, Leighly et al. 1997, 
Shapovalova et al. 2001, 2010, Sergeev et al. 2002, Peterson et al. 2004, 
Pronik \& Sergeev 2007, Sambruna et al. 2009, Jovanovic et al. 2010). 
Based on the reverberation mapping technique (Blandford \& Mckee 1982) and 
virialization method (Peterson et al. 2004, Collin et al. 2006, 
Peterson \& Bentz 2006, Peterson 2010), size of BLRs (distance between 
central black hole and broad line emission gas clouds) and central virial 
BH masses of 3C390.3 have been well determined (Peterson et al. 2004, 
Onken et al. 2004, Kaspi et al. 2005, Bentz et al. 2006, 2009, Brandon \& 
Bechtold 2007). Besides the common BH masses and size of BLRs, through 
the theoretical accretion disk model applied for observed double-peaked broad 
emission lines, some basic disk parameters of supposed theoretical 
disk-like BLRs of 3C390.3 have also been measured (Eracleous \& 
Halpern 1994), such as the inclination angle of accretion disk, the inner 
and outer radius of BLRs in accretion disk etc.. However, only due to 
the best fitted results for double-peaked balmer lines by accretion 
disk model, we can not firmly confirm that the double-peaked broad 
emission lines are exactly coming from disk-like BLRs in accretion disk 
around central black hole for 3C390.3.

    Besides the best fitted results for double-peaked broad emission lines 
by theoretical models, based on the pioneer work by Blandford \& Mckee (1982), 
geometrical structures of BLRs can be mathematically structured by 
so-called transfer function in the reverberation mapping technique through 
some special mathematical methods (such as the Maximum Entropy Method, 
Narayan \& Nityananda, 1986) applied to the properties of long-period 
observational varying both continuum emission and broad line emission 
(Peterson et al. 1993, 1994, Horne et al. 1991, Goad et al. 1993, 
Wanders \& Horne 1994, Pijpers \& Wanders 1994, Krolik 1994, Winge et 
al. 1995, Bentz et al. 2010).  However, it is unfortunate that there should 
be NOT unique solution to the so-called transfer function used to 
determine the structures of BLRs of AGN, due to the less complete 
information about variations of continuum and broad emission lines (Maoz 
1996). Thus, even based on the solution of transfer function, the supposed 
theoretical and mathematical geometrical structures of BLRs of AGN 
can NOT still be firmly confirmed.    

    In the previous remarkable work about structures of BLRs of AGN, the 
main objective is to determine the structures of BLRs through observational 
properties/characters. However, we think it will be also very interesting 
to check whether the observational properties of long-period variations of 
broad line emission and/or continuum emission could be better reproduced 
through the reverberation mapping technique, based on the expected structures 
of BLRs supposed by theoretical model, which is the main objective of 
our paper. 

  The structures of our paper are as follows. Section 2 gives some 
information about observed spectra of 3C390.3 around 1995 and how to 
find theoretical disk parameters for the supposed disk-like BLRs through 
the best fitted results for broad double-peaked balmer lines by theoretical 
elliptical accretion disk model (Eracleous et al. 1995). Section 3 shows 
the results through the reverberation mapping technique under the disk-like 
BLRs for 3C390.3. Finally section 4 gives some detailed discussions and 
conclusion. The cosmological parameters 
$H_{0}=70{\rm km\cdot s}^{-1}{\rm Mpc}^{-1}$, 
$\Omega_{\Lambda}=0.7$ and $\Omega_{m}=0.3$ have been adopted here.

\section{Fitted Results for Spectra of 3C390.3 Around 1995}

   As one well mapping double-peaked broad line AGN, 3C390.3 (z=0.056) is 
one target included in the project of AGN WATCH 
(http://www.astronomy.ohio-state.edu/\~{}agnwatch/), which is a consortium 
of astronomers who have studied the inner structure of AGN through 
continuum and emission-line variability. From the AGN WATCH project, 
we can collect 133 spectra of 3C390.3 in optical band observed around 1995 
by different instruments in different observatories. Meanwhile the 
light curves of continuum and broad emission lines after corrections of some 
necessary contamination are also collected from the website. The detailed 
descriptions about the instruments and techniques for the observed spectra 
can be found in Dietrich et al. (1998). Here, we do not describe the 
information any more. We mainly focus on how to determine the disk parameters 
by theoretical accretion disk model for 3C390.3. 

    In order to obtain reliable disk parameters for disk-like BLRs of 3C390.3 
supposed by the theoretical accretion disk model applied to fit double-peaked 
broad emission lines, we prefer to use double-peaked broad H$\alpha$ rather 
than double-peaked broad H$\beta$, because of the more apparent double 
peaks of broad H$\alpha$ (one peak of double-peaked broad H$\beta$ 
is mixed by [OIII]$\lambda4959, 5007\AA$ doublet) and more stronger broad 
H$\alpha$ than broad H$\beta$. Thus we mainly focus on the 66 out of the 
133 observed spectra with reliable broad H$\alpha$ within rest wavelength 
from 6200$\AA$ to 7300$\AA$.

   There are so far several kinds of accretion disk models which can be 
applied for double-peaked emitters, circular with/without spiral arms  
accretion disk model(Chen et al. 1989, Chen \& Halpern 1989, 
Hartnoll \& Blackman 2002), elliptical accretion disk model (Eracleous 
et al. 1995), warped accretion disk model (Hartnoll \& Blackman 2000), 
stochastically perturbed accretion disk model (Flohic \& Eracleous 2008) 
etc.. In this paper, the elliptical accretion disk model (Eracleous et 
al. 1995) is preferred, because the model can explain most of the 
observational spectral features of broad double-peaked emission lines 
(especially features for extended asymmetric line wings) with less number 
of necessary model parameters. Furthermore, the most part 
of flux density of broad double-peaked line emission is from disk-like BLRs 
into accretion disk, the existence of arms (Hartnoll \& Blackman 2002) 
and/or warped structures (Hartnoll \& Blackman 2000) and/or bright spots 
(Flohic \& Eracleous 2008) are mainly applied for subtle structures of 
double-peaked line profiles (such as some cusps around the peaks etc.), 
which have few effects on the results based on Cross Correlation Function 
in reverberation mapping technique. The detailed description of the 
elliptical accretion disk model with seven model parameters can be found 
in Eracleous et al. (1995). The seven necessary parameters are inner radius 
$r_0$, out radius $r_1$, eccentricity of elliptical rings $e$, local broadening 
velocity $\sigma$, inclination angle of disk-like BLRs $i$, line emissivity 
slope ($f_r\propto r^{-q}$) and orientation angle of elliptical rings $\phi_0$. 
Through the Levenberg-Marquardt least-squares minimization method applied 
to fit the double-peaked broad H$\alpha$ by elliptical accretion disk model 
as what we have done for the x-shaped radio source SDSS J1130+0058 
(Zhang et al. 2007), the seven disk parameters in elliptical accretion disk 
model can be well determined. The procedures to determine the disk 
parameters for BLRs of 3C390.3 are as follows.

\begin{figure*}
\centering\includegraphics[height = 16cm,width = 12cm]{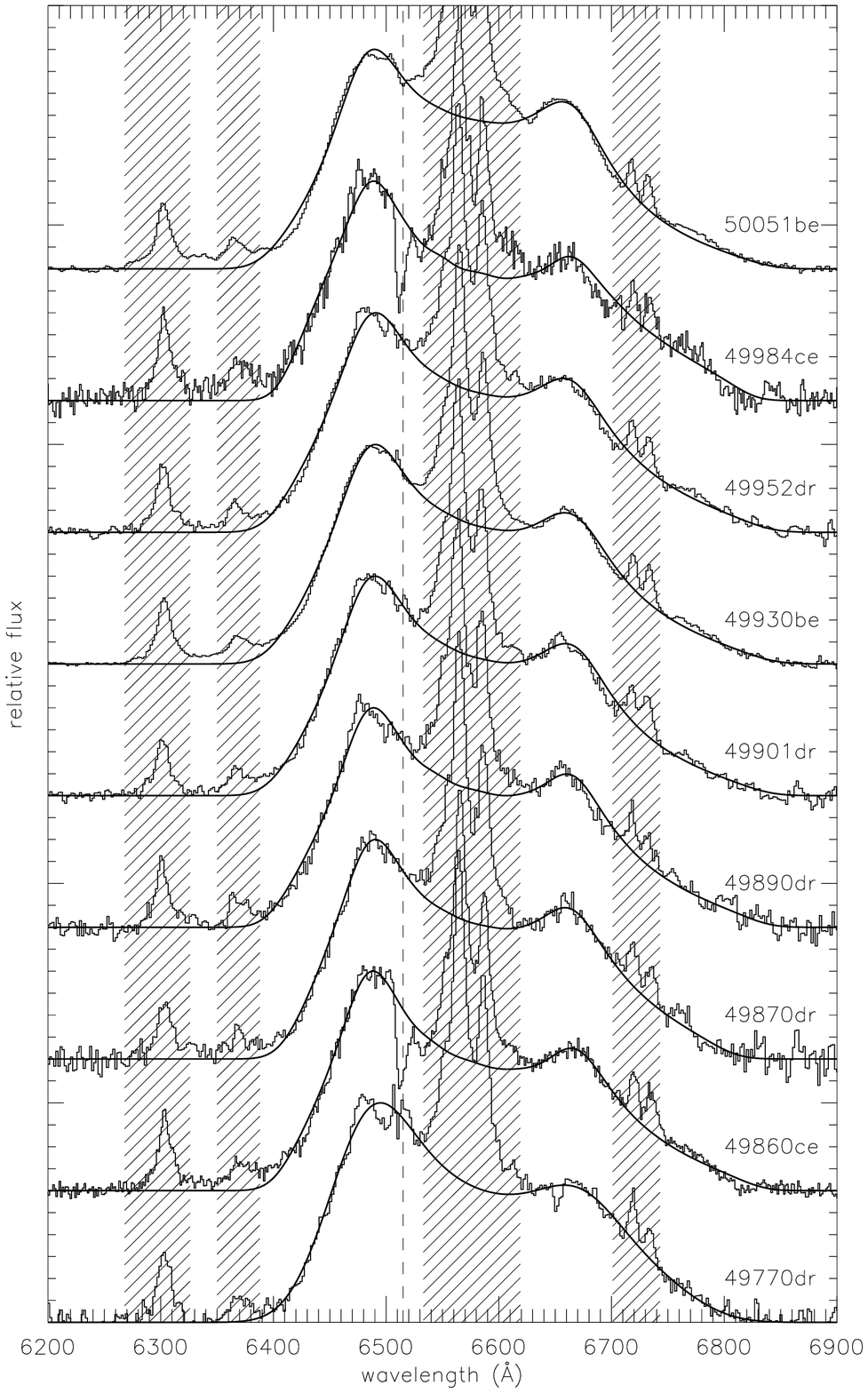}
\caption{Best fitted results for double-peaked broad H$\alpha$ in rest 
wavelength by elliptical accretion disk model. Thin solid line represents 
the observed spectrum, thick solid line represents the best fitted results 
for double-peaked broad H$\alpha$. The vertical dashed line shows the 
position for unexpected absorption features around 6500$\AA$ for two 
examples marked with '49860ce' and '49984ce'. The observed MJD date of 
each spectrum is shown in the right side of the figure. Shadow areas 
represent the ranges for narrow emission lines.}
\label{sps}
\end{figure*}

\begin{figure*}
\centering\includegraphics[height = 16cm,width = 12cm]{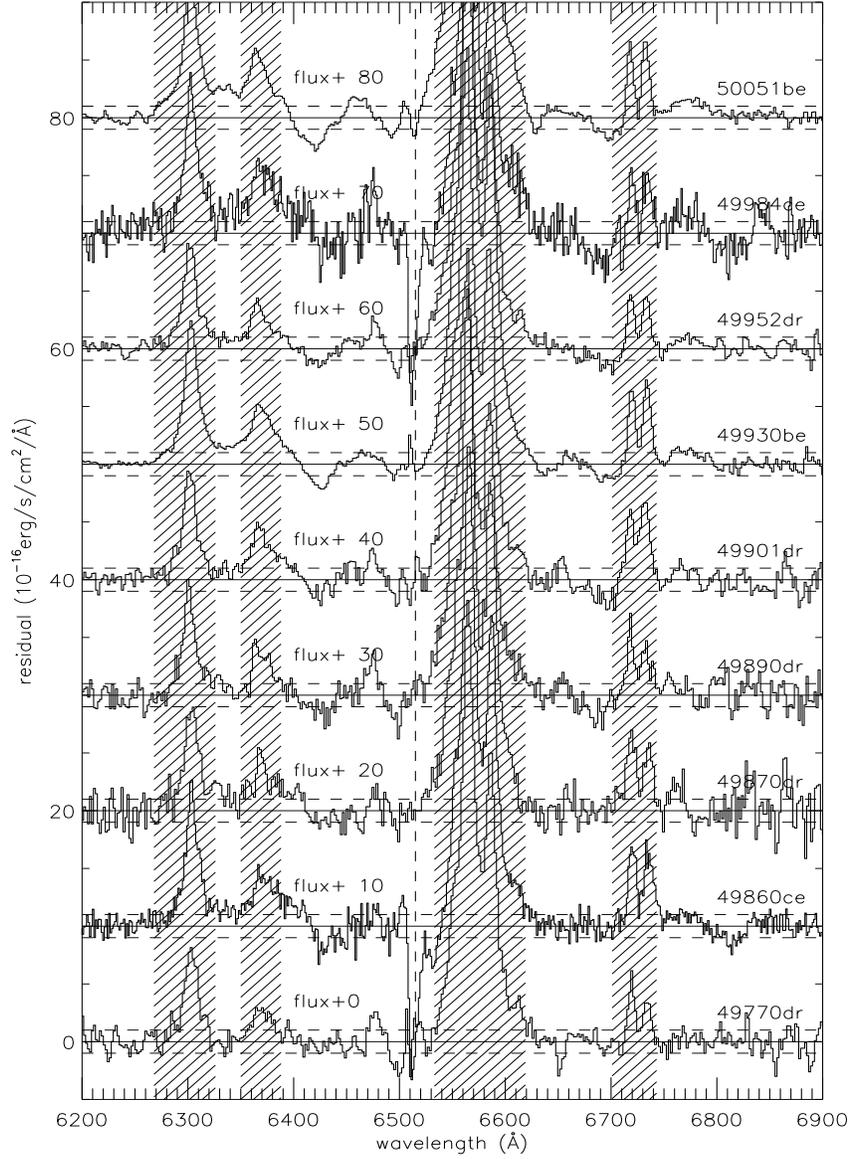}
\caption{The residuals ($y_{obs} - y_{model}$, observed data minus expected 
model data) for the shown examples in Figure \ref{sps}. Shadow areas 
represent the ranges for narrow emission lines, each double horizontal dashed lines 
represent the range of [$f_0-1$,$f_0+1$], where $f_0=0, 10, 20 ... 80$ 
representing the zero point for each spectrum are shown as solid horizontal 
lines in the figure.}
\label{res}
\end{figure*}

    First and foremost, in order to check the results from reverberation 
mapping technique under the structures of BLRs supposed by the accretion 
disk model, we should measure reliable theoretical disk parameters for 
disk-like BLRs of 3C390.3. Before starting to fit the observed broad 
double-peaked H$\alpha$, we first check the spectra of 3C390.3 by eye, 
and find that the spectra marked with 'ce' (corresponding observatory can 
be found in Table 1 in Dietrich et al. 1998) have some unexpected absorption 
features around 6500$\AA$ in rest wavelength. The weird absorption features 
can not be found in the other spectra observed by instruments in other 
observatories, which should bring some larger uncertainties in measured disk 
parameters when broad H$\alpha$ is fitted by accretion disk model. Thus the 
absorption features (rest wavelength from 6507$\AA$ to 6525$\AA$) in the 24 
spectra marked with 'ce' should be rejected, when the broad H$\alpha$ are fitted 
by the elliptical accretion disk model for 3C390.3. Besides the weird 
absorption features in the 24 spectra marked with 'ce', in order to reduce 
the effects of narrow emission lines ([OI]$\lambda6300, 6363\AA$, 
[NII]$\lambda6548, 6583\AA$, [SII]$\lambda6716, 6731\AA$ and narrow 
H$\alpha$) in all the spectra, the parts of narrow emission lines are 
also masked when broad H$\alpha$ are fitted,  rest wavelength from 
6268$\AA$ to 6326$\AA$ for [OI]$\lambda6300\AA$, rest wavelength from 
6350$\AA$ to 6388$\AA$ for [OI]$\lambda6363\AA$, rest wavelength from 
6533$\AA$ to 6619$\AA$ for [NII]$\lambda6548, 6583\AA$ and narrow 
H$\alpha$, rest wavelength from 6703$\AA$ to 6743$\AA$ for 
[SII]$\lambda6716, 6731\AA$ (the grey shadows shown in Figure \ref{sps}, 
Figure \ref{res} and Figure~\ref{unique}). In other words, only the double-peaked broad 
component of H$\alpha$ without weird absorption features and without narrow 
emission lines are fitted. After the prepared work, the double-peaked 
broad H$\alpha$ are fitted two times as follows.

    Firstly, the 39 spectra with high quality (more than 280 reliable data 
points within rest wavelength range from 6160$\AA$ to 6920$\AA$) and without 
unexpected absorption features around 6500$\AA$ are fitted by the elliptical 
accretion disk model. Then through the measured disk parameters, the two 
parameters of eccentricity and inclination angle are firstly determined for 
the 39 spectra, because eccentricity and inclination angle should not be 
changed with passage of time. The mean values of the two parameters for 
the 39 spectra are accepted as the reliable values for the two parameter. 
Then all the 66 spectra are fitted again by the accretion disk model with 
the accepted constant inclination angle and eccentricity. Figure \ref{sps} 
shows the best fitted results for some selected broad H$\alpha$ by elliptical 
accretion disk model with MJD from 49770 to 50051. In the figure, two 
spectra with unexpected absorption features (marked with '49860ce' and 
'49984ce') are also shown, the shadow areas represent the masked ranges 
for narrow emission lines. Figure \ref{res} shows the corresponding residuals 
($y_{obs} - y_{model}$, observed data minus expected model data) for the 
shown examples in Figure \ref{sps}. In the figure, shadow areas represent the 
narrow emission lines, each double horizontal dashed lines represent the 
range of [$f_0-1$,$f_0+1$], where $f_0=0, 10, 20 ... 80$ representing the 
zero point for each spectrum are shown as solid horizontal line in the 
figure. The results shown in Figure \ref{res} and in Figure \ref{sps} 
indicate that the elliptical disk model is better for double-peaked broad 
emission lines of 3C390.3, and further indicate that the probable existing 
hot spots and/or warped structures have few effects on the measured flux 
density of broad H$\alpha$ and few effects on our final results and 
conclusion, i.e., the hot spots and/or ward structures are weak for 
observed spectra of 3C390.3 around 1995, and elliptical accretion disk 
model is efficient and sufficient for 3C390.3 around 1995. 

\begin{figure*}
\centering\includegraphics[height = 12cm,width = 12cm]{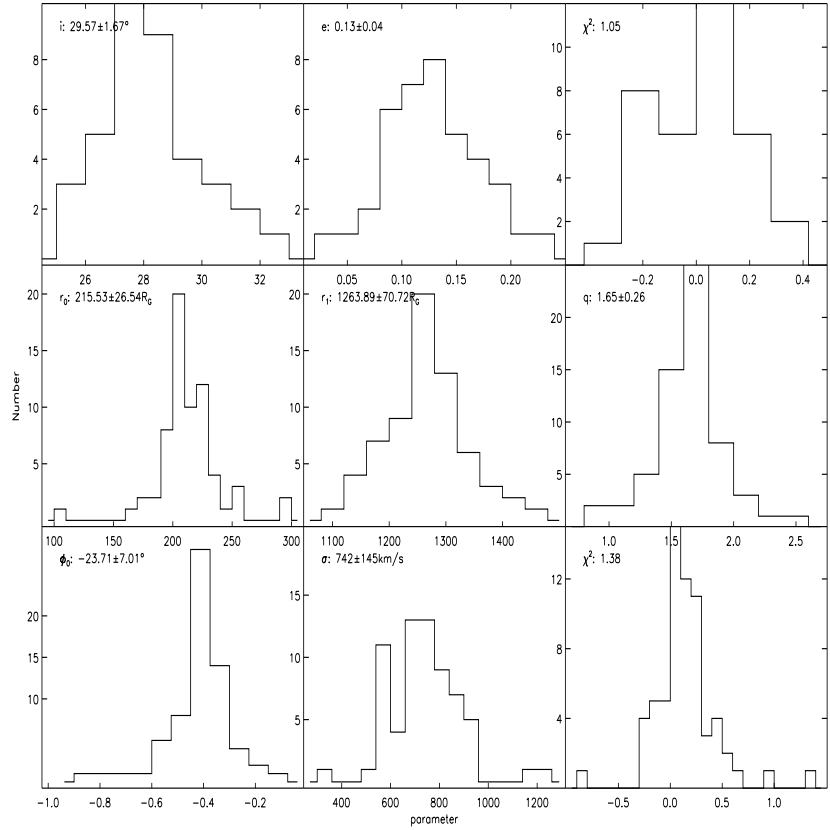}
\caption{Distributions of the disk parameters of disk-like BLRs for 3C390.3. 
The top three subfigures show the distributions of eccentricity, 
inclination angle and the parameter of $\log(\chi^2)$ for the 39 spectra 
with high quality and without unexpected absorption features around 6500$\AA$. 
Then the other six subfigures show the distributions of the other parameters 
for all the 66 spectra, inner radius $r_0$, outer radius $r_1$, 
emissivity slope $q$, orientation angle $\phi_0$, local broadening 
velocity $\sigma$ and the parameter of $\log(\chi^2)$. 
The mean value of each parameter we accepted is shown in each subfigure.
}
\label{par_dis}
\end{figure*}

   Figure \ref{par_dis} shows the distributions of the seven disk 
parameters, and the distributions of the parameter of $\log(\chi^2)$ (the value 
of the summed squared residuals divided by the degree of freedom for the
returned model parameter values, which is one parameter as the residual 
to determine whether the theoretical model is preferred) for the 39 high 
quality spectra in upper three panels and for all the 66 spectra in 
the other six panels. The final accepted disk parameters are: inner 
radius $r_0=216\pm26 R_G$, outer radius $r_1=1263\pm70 R_G$, eccentricity 
$e=0.13\pm0.04$, inclination angle $i=29.57\pm1.67\degr$, orientation 
angle $\phi_0=-23\pm7\degr$, emissivity power slope 
$q=1.65\pm0.26$ ($f(r)\propto r^{-q}$) and local broadening 
velocity $\sigma=742\pm145{\rm km/s}$. Actually, we should note that 
some of the theoretical disk parameters, $r_0$, $r_1$, $q$, $\phi_0$, $\sigma$, 
should depend on the strength of continuum emission, in other words, the parameters 
are time dependent. However, there need to be long-term progressive changes in 
the ionizing flux to change these variables. Moreover, from the results shown in 
Figure~\ref{par_dis}, we can find that there are tiny variations for these parameters, 
which indicates that the mean values for the theoretical disk parameters can be used 
to trace the actual physical disk parameters for 3C390.3 around 1995.

    Before proceeding further, we compare our measured disk parameters 
with the previous results reported in the literature. In the sample of 
double-peaked emitters from radio galaxies, Eracleous \& Halpern (1994) 
gave the results about disk parameters for BLRs through circular accretion 
disk model (Chen et al. 1989) for double-peaked broad H$\alpha$ observed in 
1988 for 3C390.3 ( simple results can also be found in Figure 6 
and in Section 4.5 in Sambruna et al. 2009), 
which are some different from our results determined through elliptical 
accretion disk model (Eracleous et al. 1995). We know that in accretion 
disk model, full width at zero intensity (FWZI) of double-peaked 
broad emission lines sensitively depends on inner radius of 
disk-like BLRs, smaller inner radius leads to broader FWZI. From the fitted 
results shown in Figure 4 in Eracluous \& Halpern (1994), we can find that 
in order to better fit the broad wings of broad H$\alpha$, the inner radius 
should be some smaller than 380$R_G$ (the one listed in Eracluous \& Halpern 
1994). Thus our inner radius $r_0=216R_G$ should be reasonable. Peak 
separation sensitively depends on the outer radius, thus there are similar 
results about outer radius in Eracluous \& Halpern (1994) and in our paper. 
In order to better fit the cusp feature around blue peak, the elliptical 
disk model should be more efficient than the totally symmetric circular 
disk model, i.e., the parameters of eccentricity and orientation angle 
should be active. The separation between peak intensity and zero intensity 
of blue (or red) part of H$\alpha$ should depends on the emissivity power 
slope. The larger distance between our inner radius and outer radius 
indicates that our emissivity slope  should be smaller than 3 (the one 
listed in Eracluous \& Halpern 1994). Thus our emissivity slope $q=1.65$ 
should be reasonable. Line width of double-peaked broad line sensitively 
depends on inclination angle, thus it is clear there are similar value about 
inclination angle $i\sim30\degr$ in our paper and in Eracluous \& Halpern 
(1994). Because the elliptical accretion disk model is more efficient for 
3C390.3 than the circular accretion disk model, the effects from local 
broadening velocity should be lower. Thus our local broadening velocity 
$\sigma\sim750{\rm km/s}$ smaller than 1900${\rm km/s}$ in Eracluous \& Halpern (1994) 
should be reasonable.

\begin{figure*}
\centering\includegraphics[height = 8cm,width = 12cm]{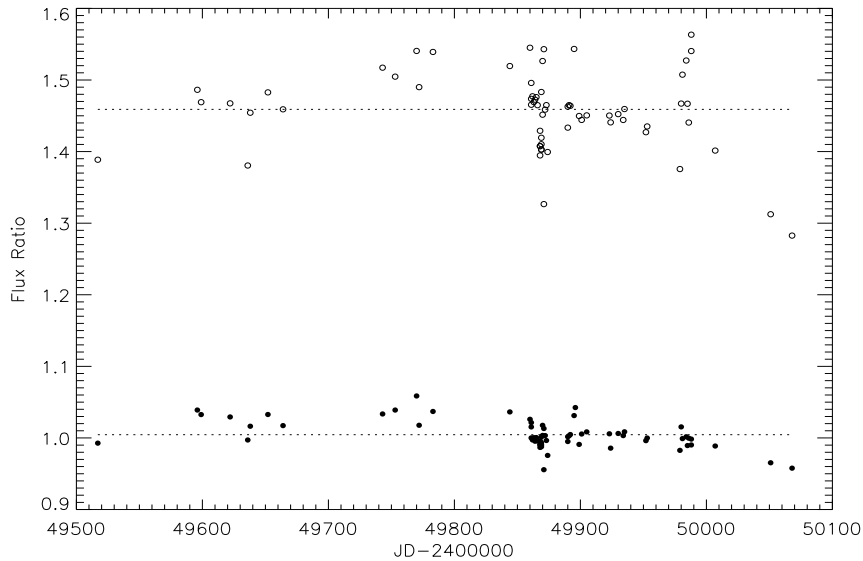}
\caption{The variations of the flux ratio of blue peak to red peak (shown as 
open circles), and the flux ratio of blue part to red part of broad 
H$\alpha$ (shown as solid circles). The dotted lines represents the 
mean values of the two kinds of flux ratios.}
\label{var}
\end{figure*}

\begin{figure*}
\centering\includegraphics[height = 16cm,width = 12cm]{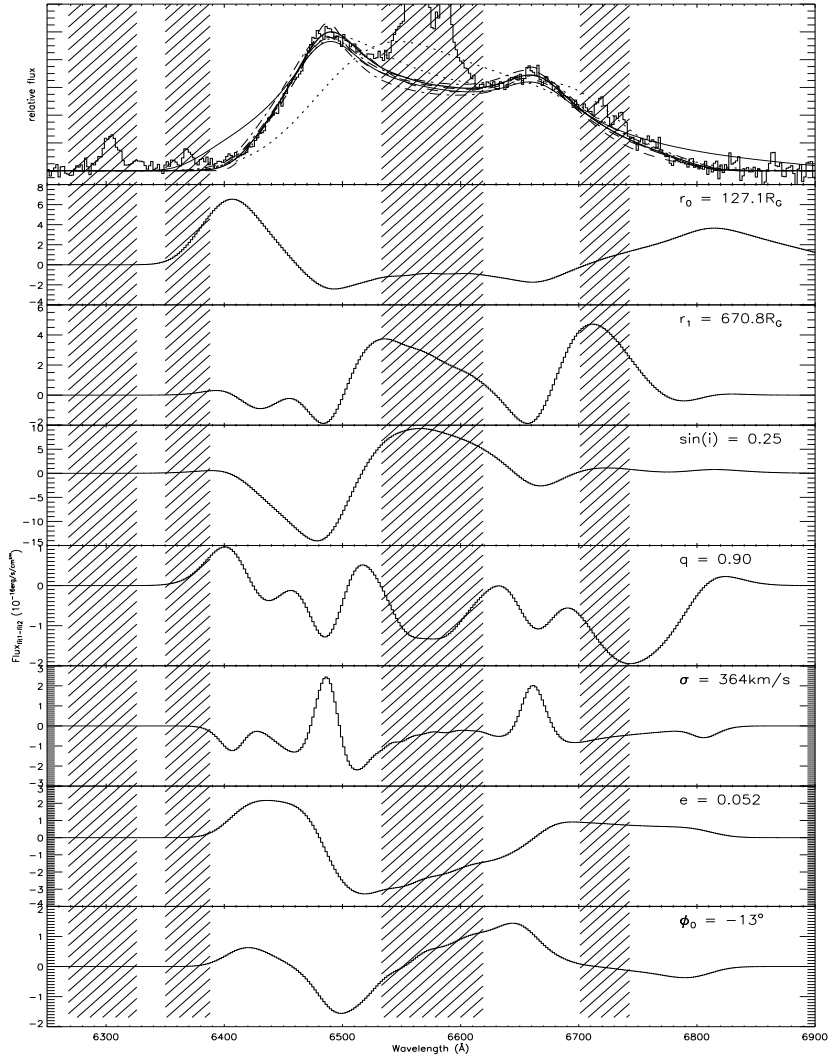}
\caption{Best fitted results for double-peaked broad H$\alpha$ with 
one fixed parameter with half of accepted value for the disk parameter. 
In the top panel, thin solid line in histogram mode represents the 
observed line profile marked with '49870dr' in AGN WATCH project, the 
thick solid line shows the best fitted results with disk parameters 
having values shown in Figure~\ref{par_dis}, thin solid line represents 
the best fitted results with disk parameter $r_0=127.1R_G$, thin 
dotted line represents the best fitted results with disk parameter 
$r_1=670.8R_G$, thick dotted line represents the best fitted results 
with disk parameter $\sin(i)=0.25$, thin dashed line represents the 
best fitted results with disk parameter $q = 0.90$, thick dashed line 
represents the best fitted results with disk parameter $\sigma = 364km/s$, 
thin dot-dashed line represents the best fitted results with disk parameter 
$e = 0.052$, thick dot-dashed line represents the best fitted results with 
disk parameter $\phi_0=-13$ in degree. The other seven panels show the corresponding 
values of $flux_{fit1-fit2}$, where $fit2$ represents the best fitted results 
with disk parameters shown in Figure~\ref{par_dis}.}
\label{unique}
\end{figure*}

    Besides the disk parameters, it is interesting to check the variations 
of line profiles of double-peaked broad H$\alpha$. As shown in Veilleux \& 
Zheng (1991), there are significant variations of relative flux ratio of 
the two peaks within the period from 1974 to 1988. Certainly, the variations 
can not be successfully explained by the elliptical accretion disk model, 
because the disk precession period of 3C390.3 based on the elliptical 
accretion disk model is about several hundreds of years as we should 
discuss in the following. However, during the period (about 500days) around 
1995 included in AGN WATCH project, the variations of flux ratio of the 
two peaks are tiny as shown in Figure \ref{var}. 
In the figure, two kinds of flux ratios are shown, the one with mean value 
of $1.01\pm0.02$ (0.02 is standard deviation for the mean value) is the 
ratio of the blue part to red part of broad H$\alpha$ divided by 
6564.61$\AA$ (the theoretical center wavelength of H$\alpha$), the other 
one with mean value of $1.46\pm0.06$ (0.06 is standard deviation) is the 
intensity ratio of the blue peak to red peak. Similar results for 3390.3 
around 1995 can also be found in Dietrich et al. (1998). The more recent 
results about the flux ratio of the two peaks of 3C390.3 from 1995 to 2007 
can be found in Shapovalova et al. (2010), there are tiny variations of 
the flux ratio of the two peaks in the 5years period as shown in Figure 13 
in Shapovalova et al. (2010). Due to the tiny variations, we believe that 
there are tiny variations for disk parameters obtained from the observed 
spectra during 1995, and that is the reason why we select the spectra 
observed around 1995. In other words, the structures of BLRs of 3C390.3 
are stable around 1995, which re-confirm that there are tiny variations 
of disk parameters shown in Figure~\ref{par_dis}.

    Last but not least, we should note the collected 66 spectra in 
optical band for 3C390.3 are the observed spectra before the 
intercaliberation method (Van Groningen \& Wanders 1992) applied to 
consider the effects from different observational instruments in different 
configurations as discussed in Dietrich et al. (1998). In this paper, we 
mainly focus on the disk parameters,  rather than the flux densities 
of broad H$\alpha$ which should be directly collected from AGN WATCH 
project with contamination being corrected. The main effect from 
intercaliberation method on the corrected line profile is the broadening 
velocity in the method. However, the broadening velocity is only about 
tens of kilometers per second when intercaliberation method is applied, 
which is much smaller than line width of broad H$\alpha$. Thus, the 
effects from intercaliberation method on the measured disk parameters 
through elliptical accretion disk model can be totally ignored. 
 
    Before to finish the section, we consider the following question 
whether there are other values for disk parameters in the disk model applied to well 
fit observed double-peaked broad H$\alpha$. In other words, the question is 
whether the solutions to the disk parameters are unique. If the answer is yes, 
there should be one set of values for the disk parameters in the following 
mathematical procedure, and we will try to determine whether our mathematical 
procedure can be applied to determine the available parameter space. If the 
answer is no, the following mathematical procedure should be simple and succinct. 
To give one precise mathematical solution to the question is very difficult. 
We consider the question as follows. The observed double-peaked broad 
H$\alpha$ is re-fitted. When procedure starts to fit line profile, one of 
the disk parameters is fixed to one value much different from the accepted 
value above (half of the accepted value for the parameter), and the other 
disk parameters with the same starting values in fitting procedure are free. 
Then the Levenberg-Marquardt least-squares minimization technique is applied 
to find the best fitted results for observed double-peaked broad H$\alpha$ 
and find the best solutions for the disk parameters. Actually, the software 
package MPFIT in Markwardt IDL Library (http://www.physics.wisc.edu/~craigm/idl/) is used 
to perform the least-squares fitting and to find the best solution. In order 
to clearly compare the fitted results for different disk parameters, the value 
$flux_{fit1-fit2}$ is calculated, where $fit2$ means the best fitted results 
for accepted disk parameters shown in Figure~\ref{par_dis} and $fit1$ means 
the best fitted results for one fixed disk parameter with half of the accepted 
value for the parameter. Figure~\ref{unique} shows the fitted results, and the 
corresponding values of $flux_{fit1-fit2}$. The apparent and large difference 
(residual to some extent) between $fit1$ and $fit2$ indicates that solutions 
to disk parameters based on accretion disk model are unique to some extent. 
Thus, in the following mathematical procedure to determine geometrical structure 
of emission line regions, there are fixed disk parameters. 

    Based on the results above, through the disk parameters, one kind 
disk-like geometrical structures of BLRs of 3C390.3 can be well structured. 
Then it is very interesting to check whether the characters of observed 
light-curves can be reproduced through the theoretical disk-like BLRs 
of 3C390.3, which is the main objective of the following section.

\section{Results about Reverberation Mapping Technique Under Elliptical 
Accretion Disk Model}

   In the section above, one kind geometric structure of BLRs of 3C390.3 
have been supposed by the theoretical elliptical accretion disk model 
(Eracleous et al. 1995). In this section, we will mainly check whether 
the supposed disk-like BLRs are valid to reproduce properties of observed 
continuum and broad line variability, through reverberation mapping technique. 

   In order to clearly build one physical geometrical structure of disk-like 
BLRs, the length of one gravitational radius ($R_G$) in physical unit should 
be firstly determined, i.e., it is necessary to determine the masses of 
central black hole of 3C390.3. Here the BH masses of 3C390.3, 
$M_{BH} = 4.68\times10^8{\rm M_{\odot}}$, is accepted. For 3C390.3, the two 
kinds of BH masses, both virial BH masses (Peterson et al. 2004) and BH 
masses from M-sigma relation (Tremaine et al. 2002, Ferrarese \& Merritt 2001, 
Gebhardt et al., 2000, G\"ultekin et al. 2009, Woo et al. 2010,  Lewis \& 
Eracleous 2006) can be found in literature, which are similar. Thus we do 
not worry about the accuracy of BH masses of 3C390.3, and accept 
$R_G\sim0.0263{\rm light-days}$ based on the BH masses. Through the disk 
parameters above, it is not difficult to build the physical geometry of the 
BLRs of 3C390.3 supposed by theoretical model. Then it is interesting to 
check the results based on the reverberation mapping technique through the 
supposed disk-like BLRs as follows.

    First and Foremost, we accept the assumptions listed in Peterson (1993) 
for reverberation mapping technique, 1): the continuum emission is from one 
central source which is much smaller than BLRs. The assumption is efficient 
for 3C390.3, although double-peaked broad line of 3C390.3 is assumed from 
accretion disk. The reported size of BLRs of 3C390.3, the time lag between 
observed optical balmer emission and observed UV X-ray emission, is about 
$20\pm8$ light-days (Dietrich et al. 1998, Bentz et al. 2009), i.e., 
$R_{BLRs}\sim400 - 1000{\rm R_G}$, through cross correlation function 
method. Based on the theoretical disk parameters for 3C390.3, the flux 
weighted mean size of BLRs is about $567{\rm R_G}$ (simply calculated by 
$(r_0+r_1)/(1.+q)$, $q\sim1.65$ is the line emissivity slope). The result 
indicates the UV emission region is much near to central black hole, which 
clearly can be treated as one point source. 2): both continuum emission 
and line emission are freely and isotopically propagating in the central 
volume, which can be confirmed by the small covering factor estimated by 
the appearance of no-absorption features around broad H$\alpha$. 3): the 
line emissions are in rapid response to ionizing continuum, which can be 
confirmed by the strong correlation between line luminosity and continuum 
luminosity found by Greene \& Ho (2005) and by the much shorter recombination 
time (about one hundred seconds for standard BLRs) of balmer emission lines 
and much longer dynamic time (about several years for standard BLRs) than 
the light travel time across BLRs, as discussed in Peterson (1993). 

\begin{figure*}
\centering\includegraphics[height = 13cm,width = 16.5cm]{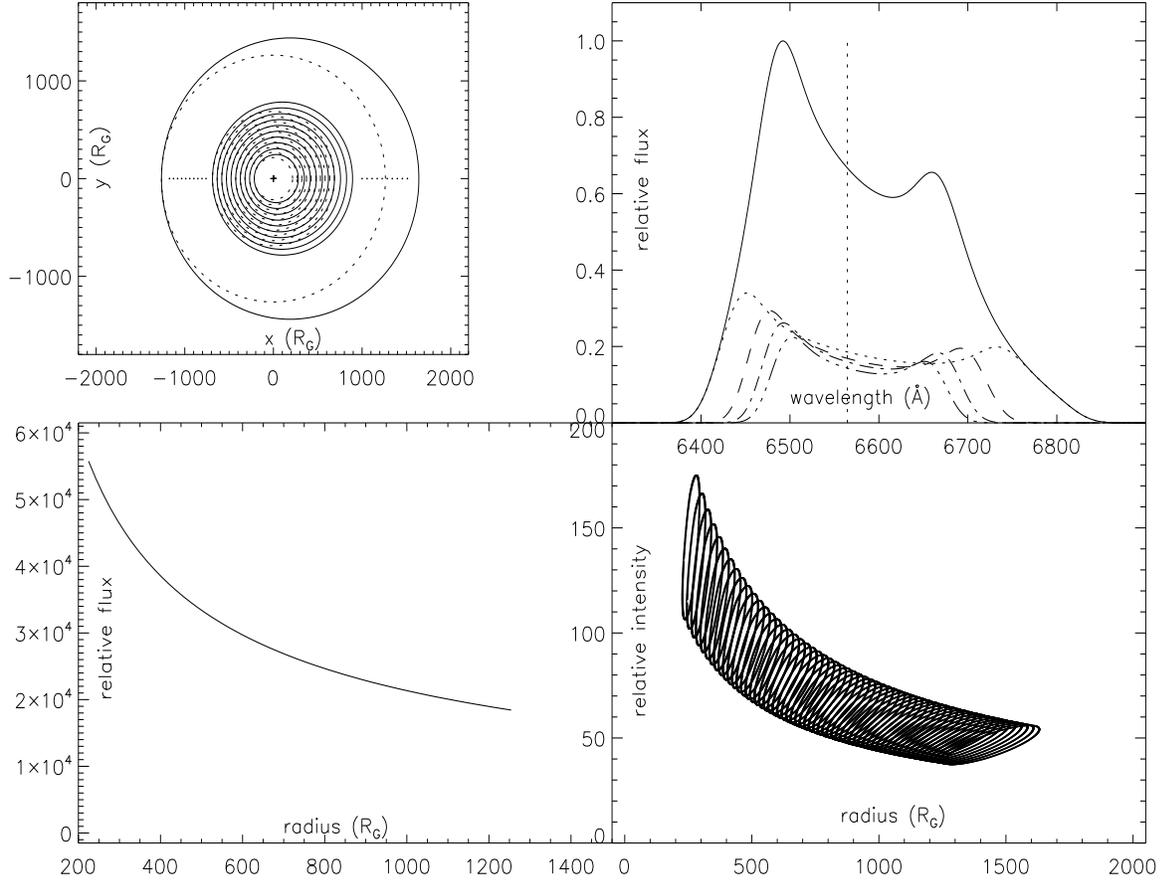}
\caption{Top left panel shows the toy structures of the elliptical disk-like 
BLRs of 3C390.3. The solid line represents the elliptical ring with 
pericenter distance  $r_0\le r_{i}\le r_1$, the innerest one with pericenter 
distance of $r_0$, the outerest one with $r_1$. The central black hole is 
located at the origin of the coordinates (one focus point of the elliptical 
ring). Dotted line represents the sphere surface of ionizing photos at one time. Because 
the BLRs are elliptical disk-like, thus at one time, the ionizing 
photos arriving in BLRs DO NOT affect all the tiny clouds located in 
one elliptical ring. Top right panel show the toy showing of the line 
intensity of broad H$\alpha$ from different bins of radius. Solid line 
represents the observed H$\alpha$, dotted line represents the line intensity 
from $\sim216R_G$ (the inner boundary) to $\sim480R_G$, dashed line 
represents the line intensity from $\sim480R_G$ to $\sim740R_G$, dot-dashed 
line represents the line intensity from $\sim740R_G$ to $\sim1000R_G$, 
double-dot-dashed line represents the line intensity from $\sim1000R_G$ 
to $1263R_G$ (the outer boundary). The vertical dotted line represents 
the center wavelength of H$\alpha$, 6564.61\AA. Bottom left panel shows 
the correlation between line intensities from different bins ($F_i(r)$) and the 
corresponding pericenter distances of the bins. Here the bins are 
created uniformly, thus $F(r)\propto r^{(\sim(1-q))} \propto r^{(-0.65)}$.
Bottom right panel shows the properties of line intensities ($f_{i,j}$) of the 
23541 data points in sixty elliptical rings with pericenter distances 
from $r_0$ to $r_1$. The bottom two panels show the properties of Equation (2).
}
\label{math}
\end{figure*}

   Besides, we structure the supposed theoretical structures of BLRs of 
3C390.3 as follows. More than $10^4$ tiny clouds (or so-called test 
particles, if we treat each tiny cloud with sufficient tiny size) are set 
in the N elliptical rings (the BLRs), each tiny cloud has its position 
(radius $r$ and orientation angle $\phi$) and line intensity ($f$). 
The position in one elliptical ring for one tiny cloudy is created by,
\begin{equation}
\begin{split}
\phi &\in [0, 2\times\pi] \\
r &= \frac{r_{\star}\times(1+e)}{1-e\times\cos(\phi)}
\end{split}
\end{equation}
where $r_0\le r_{\star}\le r_1$ and $e$ are the pericenter distance and 
eccentricity of the disk-like BLRs of 3C390.3, $r$ is the distance from 
the tiny cloud to the central black hole. From inner boundary $r_0$ to 
outer boundary $r_1$, the pericenter distance for the elliptical rings is 
evenly separated into N bins, 
$\log(r_0)\le \log(r_{\star,i}(i=1\dots N))\le \log(r_1)$ 
with step of $(\log(r_1)-\log(r_0))/(N-1)$. Because most part of line 
emission is from inner area of BLRs, thus the parameter of logarithm 
of pericenter distance ($\log(r)$) is evenly separated, rather than the 
parameter of pericenter distance ($r$), which provides more areas in inner 
part of BLRs. After that, the orientation angle $\phi$ is evenly separated 
into M bins, $0\le \phi_{j}(j=1\dots M)\le 2\times\pi$ with step of 
$2\times\pi/(M-1)$. Based on the $r_{\star,i}$ and $\phi_j$, the supposed 
theoretical disk-like BLRs can be separated into $(N-1)\times (M-1)$ tiny 
areas ($A_{i,j}(i=1\dots N, j=1\dots M)$). As long as N is larger enough, 
the ionizing photo propagating time through each tiny area can be ignored, 
and each tiny cloud can be simply treated as one point source. The line 
intensity from each area $f_{i,j}(i=1\dots N, j=1\dots M)$ can be directly 
calculated by elliptical accretion disk model, and the total line 
intensity $F$ is the sum of $f_{i,j}$,
\begin{equation}
\begin{split}
&f_{i,j} = \int\limits_{r_{\star,i}}\limits^{r_{\star,i+1}}\int\limits_{\phi_j}\limits^{\phi_{j+1}}
        H(model)drd\phi \propto P(r_{i}, \phi_{j})\\
&\sum\limits_{j} f_{i,j} = F_i \propto r_{\star,i}^{q_{\star}}\\
&\sum\limits_{i}\sum\limits_{j} f_{i,j} = F
\end{split}
\end{equation}
where $H(model)$ is the integrand function used for accretion disk model 
as shown in Eracleous et al. (1995), $q_{\star}$ is different from the 
emissivity slope in elliptical accretion disk model (however if the 
radius of each bin is uniformly created, then $q_{\star} \sim 1.0 - q$ as 
shown in Figure \ref{math}), $P(r_{i}, \phi_{j})$ is the function 
depending on radius and orientation angle to calculate the line intensity 
of each tiny cloud in the $ith$ elliptical ring which can also be found 
in Figure \ref{math}, $F_{i}$ and $F$ mean the line intensity from the 
$ith$ elliptical ring and the total line intensity from total area. Here, 
we select $N = 60$ and $M=400$. $N=60$ confirms that the light travel 
time from the $i$ bin to $i+1$ bin is much less than 1 day (especially for the 
bins in inner part of BLRs), the standard date separation of our input 
light curve of continuum emission. Furthermore, we should note that 
in our procedure, the listed disk parameters above are taken as fixed values, 
based on the following considerations. On the one hand, there are tiny 
variations of line profiles of double-peaked H$\alpha$ observed around 1995. 
Based on the observed H$\alpha$ selected rom AGNWATCH, it is very difficult 
to find one reliable correlation between line profile variability and disk 
parameters (especially orientation angle $\phi$) through theoretical disk model. 
In other words, there is no information about the function of $\phi(t)$ where 
$t$ is date and time. On the other hand, the fixed disk parameters through tiny 
varied line profiles around 1995 lead to much simple mathematical procedure.

\begin{figure*}
\centering\includegraphics[height = 7.5cm,width = 12cm]{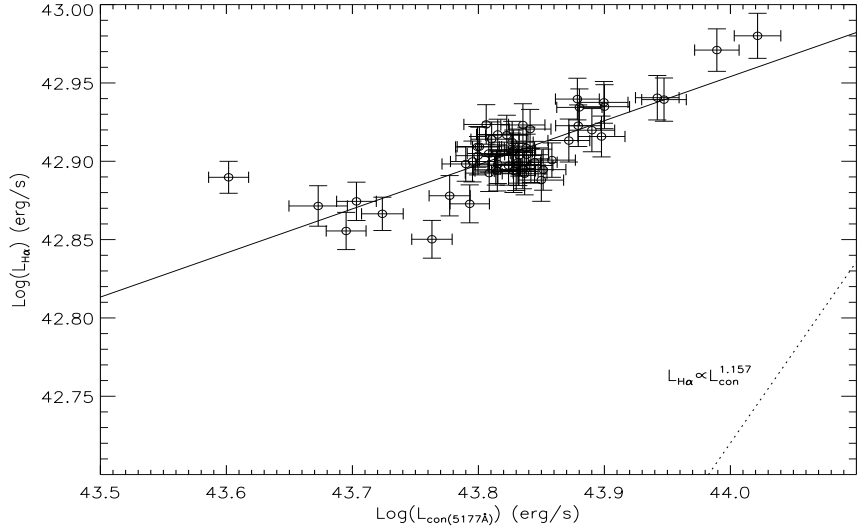}
\caption{The correlation between luminosity of broad H$\alpha$ and 
continuum luminosity at 5177$\AA$ for 66 observed spectra around 1995. Solid 
line represents the best fitted result, 
$L_{H\alpha}\propto(L_{con(5177\AA)})^{0.265}$. 
The dotted line is the one found by Greene \& Ho(2005).}
\label{conha}
\end{figure*}

   Figure \ref{math} shows the procedures above to build the structures of BLRs 
of 3C390.3 based on the disk parameters,  including the properties of 
equation (2). Once the tiny cloud meet the 
ionizing photo from central source, the line intensity of the tiny cloud 
is changed immediately as,
\begin{equation}
f_{i,j}(t) \propto f_{i,j}(t-1)\times (\frac{con_{i,j}(t)}{con_{i,j}(t-1)})^{\alpha}
\end{equation}
where $t$ is the date with uniform separation of 1 day as discussed below, 
$\alpha$ is slope of the the corresponding correlation between luminosity of 
H$\alpha$ and continuum luminosity, $f_{i,j}$ and $con_{i,j}$ mean the line 
intensity from area $A(i,j)$ and arriving continuum emission for area  
$A(i,j)$. For AGN, $\alpha$ is the about 1 for QSOs in Greene \& Ho (2005), 
and for low luminosity AGN in Zhang et al. (2008). However, $\alpha\sim1$ is 
not preferred for 3C390.3, Figure \ref{conha} shows the correlation between 
continuum luminosity at 5177$\AA$ and luminosity of broad H$\alpha$ for the 
spectra observed around 1995. The values of luminosity of broad H$\alpha$ 
and continuum luminosity at 5177$\AA$ are collected from AGN WATCH, the 
contamination has been corrected. The spearman rank correlation coefficient 
is about 0.66 with $P_{null}\sim10^{-9}$. Through the Levenberg-Marquardt 
least-squares minimization technique, the best weighted fitted result for 
the correlation can be calculated,
\begin{equation}
L_{H\alpha} = 8.92\times10^{42}(\frac{L_{con(5177\AA)}}{10^{44}{\rm erg/s}})^{0.26\pm0.03} {\rm erg/s}
\end{equation}
Thus, we accept $\alpha\sim0.26$, not $\sim1$. Actually, the different index 
of line-continuum luminosity correlation is due to the contamination of narrow lines 
around H$\alpha$. As shown in Dietrich et al. (1998), the flux density of broad H$\alpha$ 
includes contributions from narrow lines around H$\alpha$, and flux densities of narrow 
emission lines are constant for 3C390.3. If we define factor $k$ as flux ratio 
of pure broad H$\alpha$ to narrow lines, we should find that points with lower luminosity of 
H$\alpha$ in Figure~\ref{conha} have smaller values of $k$, points with larger luminosity of 
H$\alpha$ in Figure~\ref{conha} have larger values of $k$. If the contributions of narrow 
emission lines are corrected, the index for the correlation shown in Figure~\ref{conha} should be 
steeper than 0.26 and near to 1.

\begin{figure*}
\centering\includegraphics[height = 8cm,width = 12cm]{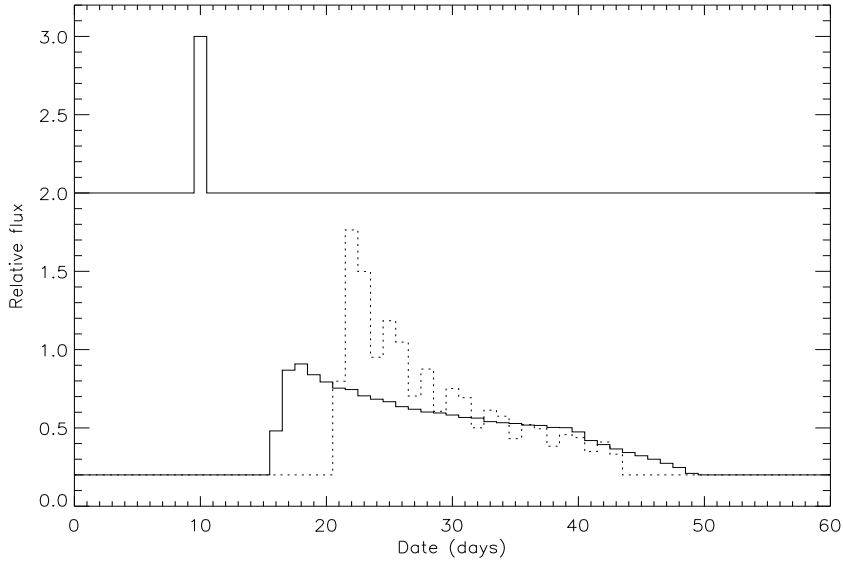}
\caption{ The response output broad H$\alpha$ based on the input 
continuum emission described by one delta function under 
the elliptical disk-like BLRs and circular disk-like BLRs for 3C390.3. The top line 
shows the continuum emission. The bottom solid line shows the corresponding 
response output broad H$\alpha$ under elliptical disk-like BLRs, and 
dotted line shows the results under circular disk-like BLRs.}
\label{example}
\end{figure*}

     Figure \ref{example} shows one simple output line curve of broad 
H$\alpha$ based on one test input light curve of continuum emission described 
by one delta function through the disk-like BLRs determined elliptical accretion 
disk model for 3C390.3. Because of the extended BLRs, the delta function is 
delayed and expanded. Light traveling from the central black hole (we 
assume that the ionizing photos are from central point) to the innerest 
boundary of the BLRs is about 5days, from the inner boundary ($\sim216R_G$) 
to the point with longest distance ($\sim1640R_G$, larger than the pericenter 
distance $1263R_G$ for the out boundary, because the BLRs are elliptical 
disk-like) from central black hole is about 39 days. Thus in the figure, it 
is clear that 5 days after the burst of continuum, the line intensity 
starts to vary, and then 39 days after the burst of continuum emission, the 
variation of line intensity ends. Because the inner part of BLRs emits most 
of the line intensity of broad H$\alpha$, thus the line intensity of total 
H$\alpha$ is changed from strong to weak with passage of time as shown in 
the figure. Furthermore, it should be interesting to check the effects 
of different geometrical structure on the response output broad H$\alpha$, although 
we have accepted there are fixed and stable disk parameters in section above. 
We accepted the disk parameters for 3C390.3 under the stochastically perturbed 
accretion disk model in Flohic \& Eracleous (2008, one circular disk part with 
bright spots), $r_0=450R_G$, $r_1=1400R_G$, $\sigma=1300{\rm km/s}$, $i=27^{o}$ and 
$q=3$, in spite of the not well fitted results for line profiles with these 
parameters (not good fitted results for broad wings). Under the model, the 
output H$\alpha$ is also shown in Figure~\ref{example}, It is clear that 
the response to input continuum described delta function is much different 
for elliptical disk structure and 
for circular disk structure.  The different response for different geometrical 
structures indicates the different structures of BLRs of AGN should be 
discriminated through the light curves of continuum emission and broad 
emission lines emission, especially when there are HOMOGENEOUS and 
COMPLETE information for the light curves. Actually, based on the observed 
light curves of 3C390.3 selected from AGN WATCH project, there is no way 
to discriminate different geometrical structures of BLRs as what we have 
shown in the following results. The results shown in the figure indicate that our 
procedures above DO work, and effects of extended disk-like BLRs on light curve of 
emission line can be clearly described by our procedures. 

\begin{figure*}
\centering\includegraphics[height = 12cm,width = 16cm]{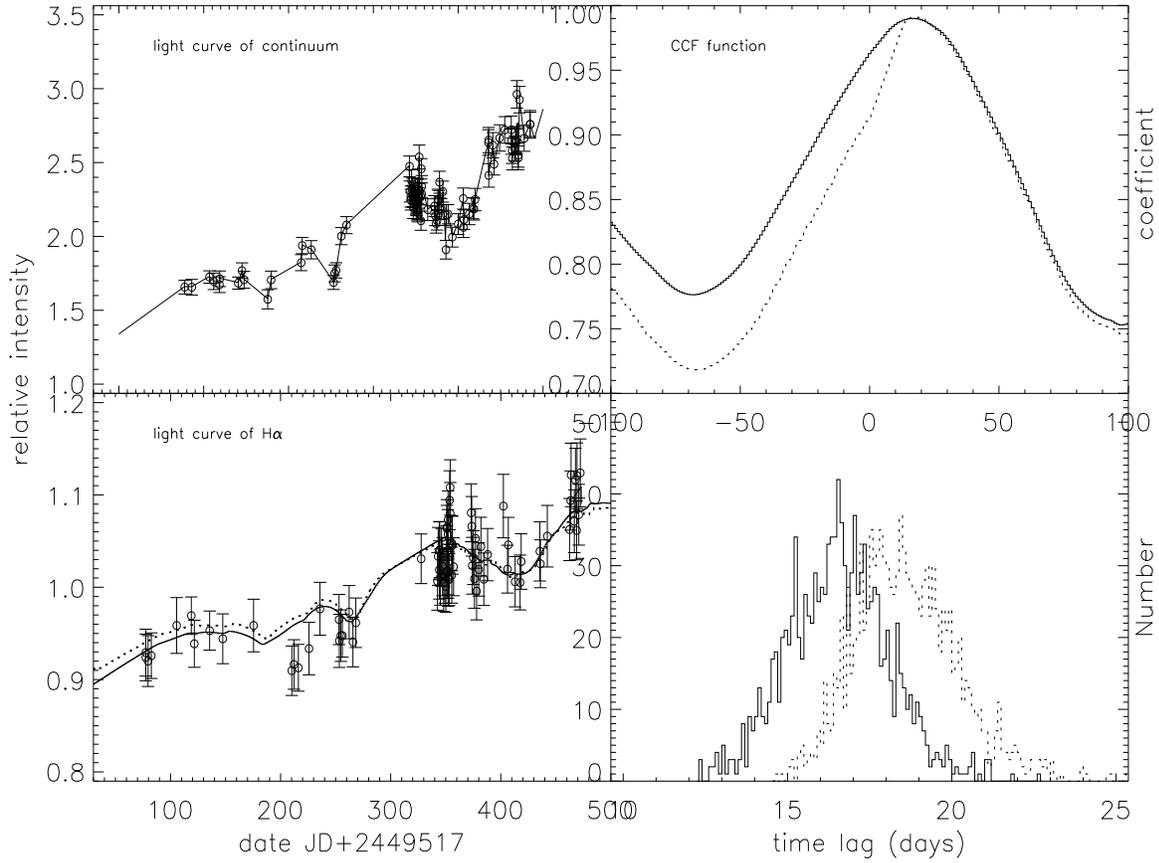}
\caption{Top left panel shows the light curve of continuum, open circles are 
the observed data points selected from AGN WATCH, solid line represents the 
input light curve of continuum with separation of 1day. Bottom left 
panel shows the light curve of broad H$\alpha$, open circles are observed 
values, solid line represents the output light curve under the elliptical 
disk-like BLRs, dotted line represents the output light curve under the 
circular disk model shown in Flohic \& Eracleous (2008). Top right panel shows 
the CCF function (maximum coefficient about 0.99) for observed light-curve of 
continuum emission and output light curve of broad H$\alpha$.  Solid line 
represents the result for elliptical accretion disk model (peak value around 
16 days), dotted line shows the result for circular disk model in Flohic \& 
Eracleous (2008) (peak value around 18 days). Bottom right shows the 
distribution of time lag between observed continuum emission and simulated 
broad H$\alpha$ emission by bootstrap method. Solid line is for elliptical 
disk model, dotted line is for circular disk model in Flohic \& Eracleous (2008).}
\label{obs}
\end{figure*}

   Last but not least, the observed light curve of continuum emission at 
$\sim5177\AA$ collected from AGN WATCH is used as the input light curve 
of continuum emission. The detailed description about the light curve can 
be found in Dietrich et al. (1998). There are 70 data points included in 
the light curve from MJD2449517 to MJD2450068. We first create one new 
light curve of continuum emission with date separation of 1day by linear 
interpolation method applied to observed light curve of continuum. Then we 
check the expected output light curve of broad H$\alpha$ under the supposed 
elliptical disk-like BLRs. The results are shown in bottom-left panel of 
Figure \ref{obs}. It is clear that the expected output light curve of 
broad H$\alpha$ is consistent with the observed light curve of broad 
H$\alpha$, the linear correlation coefficient for the correlation 
between output light curve of H$\alpha$ and observed light curve of H$\alpha$ 
is about 0.85 with $P_{null}\sim0$. Then through the input light curve of 
continuum and the output corresponding light curve of broad H$\alpha$, it 
is interesting to estimate the size of BLRs by so-called cross correlation 
function (CCF). Here, the common interpolated cross-correlation function 
(ICCF) (Gaskell \& Sparke 1986, Gaskell \& Peterson 1987, Peterson 1993) 
is applied to quantify time lag between continuum emission 
and broad lines emission. We do not consider the z-transfer discrete 
correlation function (ZDCF, Alexander 1997, Edelson \& Krolik 1988, 
White \& Peterson 1994) any more, because results from ZDCF are excellent 
agreement with the results from ICCF (Peterson et al. 1991, 1992, 2004, 
Kaspi et al. 2000, Bentz et al. 2010). Furthermore, the corresponding 
results based on the disk parameters shown in Flohic \& Eracleous (2008) 
are also calculated, and shown in Figure~\ref{obs}. It is clear that 
although the example shown in Figure~\ref{example} indicates 
the information from light curves can be used to confirm which structure 
should be more preferred, there is no way to find enough evidence to 
determine which structure (elliptical disk-like BLRs or circular disk-like 
BLRs) is preferred, due to the incomplete and in-homogeneous information 
provided from the observed light curves. Thus, we mainly compare the 
calculated time lags between continuum emission and broad line emission, 
under the two structures.

    Figure \ref{obs} shows the CCF results from input light curve of 
continuum and the output light curve of broad H$\alpha$, the distribution 
of the size of BLRs through bootstrap method (a common Monte-Carlo method to 
estimate the uncertainty of parameter, Press et al. 1992, Peterson et al. 1998). 
Detailed descriptions about CCF method to estimate size of BLRs (time 
lag between line emission and continuum emission) and bootstrap method 
to estimate uncertainties of measured time lag can be found in Peterson 
(1993). From the results shown in the figure, we can find that the 
estimated size of BLRs through the input light curve of continuum emission 
and output light curve of broad H$\alpha$ under elliptical accretion disk 
model is also well consistent with the one estimated by the observed light 
curves of continuum emission and broad line emission, 
$R_{BLRs}\sim16.8\pm2.1{\rm light-days}$. 
We should note that the size of BLRs is the distance from broad H$\alpha$ emission 
line region to optical continuum emission region, thus the value is some smaller 
than $\sim20{\rm light-days}$ estimated by the CCF function from light curve 
of broad H$\alpha$ and light curve of x-ray/UV emission shown in Dietrich 
et al. (1998), Bentz et al. (2009) etc., because there are probable 5days 
lag between optical continuum emission and x-ray emission (Dietrich et 
al. 1998). Certainly, we also calculate the results based on the circular 
disk model shown in Flohic \& Eracleous (2008), which should be some different 
from the results for elliptical accretion disk $R_{BLRs}\sim18.4\pm2.1{\rm light-days}$. 
However, there is no enough evidence to against the circular disk-like BLRs for 3C390.3, 
besides the best fitted results for observed line profiles. If there should homogeneous 
and complete observed light curves, the clear final decision should be given.

\begin{figure*}
\centering\includegraphics[height = 7cm,width = 10cm]{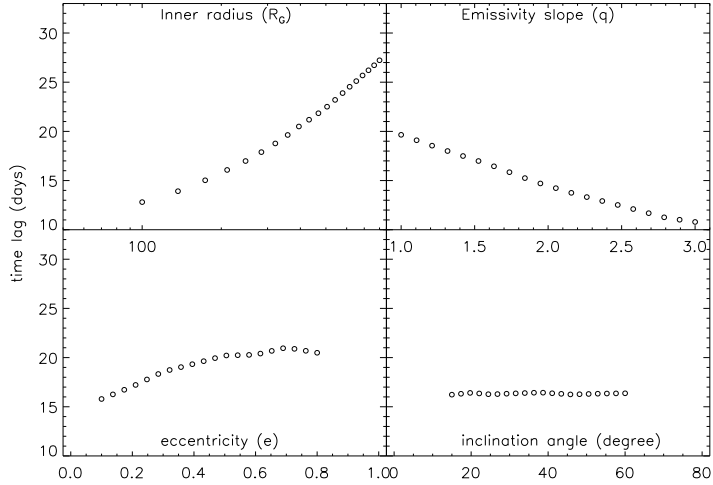}
\caption{The effects of disk parameters on the theoretically measured size 
of BLRs. Top-left panel shows the effects of inner radius, top-right panel 
shows the effects of emissivity power, bottom-left panel shows the effects 
of eccentricity, bottom-right panel shows the effects of inclination angle.}
\label{par_ccf}
\end{figure*}

   In order to further confirm that the disk-like BLRs determined by 
elliptical accretion disk model is reliable for double-peaked emitter 
3C390.3, we also check the CCF results for disk-like BLRs with different 
disk parameters, in spite of the best fitted results for observed 
double-peaked broad balmer line. Figure~\ref{par_ccf} shows the effects 
of disk parameters on the measured size of BLRs through input continuum 
emission and output H$\alpha$ emission based on the elliptical accretion 
disk model as what we have done above. It is clear that inner radius and 
emissivity power are the two main parameters which have apparent effects 
on the theoretically measured size of BLRs of 3C390.3. In the figure, we 
only show the effects from inner radius, emissivity power, eccentricity and 
inclination angle of disk-like BLRs, because there are few effects from the 
other disk parameters. Certainly, more reliable evidence to determine and 
confirm model parameters  should depend on future more detailed and homogeneous 
observed light curves of continuum emission and broad line emission. 
The results indicate that the measured disk parameters for BLRs of 3C390.3 
are reliable to some extent, and different geometrical structures of BLRs 
based on different disk parameters should lead to different theoretically 
measured size of BLRs from size of BLRs through observational results.  

   The results above indicate that the elliptical accretion disk model can 
be applied to best fit the observed spectra, i.e., the model can provide 
fine velocity structures for BLRs of 3C390.3. Furthermore, the elliptical 
accretion disk model can provide fine geometrical structures which can be 
applied to reproduce characters of observed light curve of broad H$\alpha$. 
Thus elliptical accretion disk model which can be applied to fit characters 
of line profiles and variations of broad double-peaked H$\alpha$ is 
appropriate to double-peaked emitter 3C390.3.

\section{Discussion and Conclusion}

   As one well-known double-peaked emitter, based on the disk parameters 
measured above, to check the expected line profile of double-peaked broad 
balmer emission line of 3C390.3 in different years/periods is very 
interesting. Although the more recent observed line profiles of double-peaked 
broad H$\alpha$ of 3C390.3 in Jan. 2007 can be found in Shapovalova et al. 
(2010), the observed spectrum in Sep. 2000 by HST STIS is collected, because 
the reduced spectrum in 2000 is public and conveniently collected from 
the website of Multimission Archive at STScI (MAST) 
(http://archive.stsci.edu/) (HST Proposal 8700, PI: Prof. Andre Martel 
in Space Telescope Science Institute). The detailed description about 
observational technique for the spectrum around 2000 by HST can be found 
in Popovic (2003). Here, we mainly check whether the observed line 
profile of broad balmer emission lines in 2000 can be reproduced through 
theoretical elliptical accretion disk model based on the disk parameters 
measured from spectra observed around 1995 for double-peaked emitter 3C390.3. 

   Before proceeding further, the precession period should be determined. 
Through the obtained disk parameters, the relativistic precession period 
of the elliptical disk-like BLRs around central black hole of 3C390.3 can 
be simply calculated (Weinberg 1972), 
\begin{equation}
\begin{split}
T_{pre} &= \frac{2\times\pi}{\delta_{\phi}} \sim 
\frac{2\times\pi}{\frac{6\pi\times G\times M_{BH}}{c^2A(1-e^2)}} \\
  & \sim 70 {\rm years}\hspace{4mm} ({\rm inner\ \ radius})\\
  & \sim 370 {\rm years}\hspace{4mm} ({\rm outer\ \ radius})
\end{split}
\end{equation} 
where 'A' is the semi-major axis length and $e$ is the eccentricity. 
Although, the precession period for outer elliptical rings is very long, 
the short precession period for inner elliptical rings should lead to some 
apparent variations of double-peaked broad emission lines from 1995 to 2000. 

\begin{figure*}
\centering\includegraphics[height = 7cm,width = 11cm]{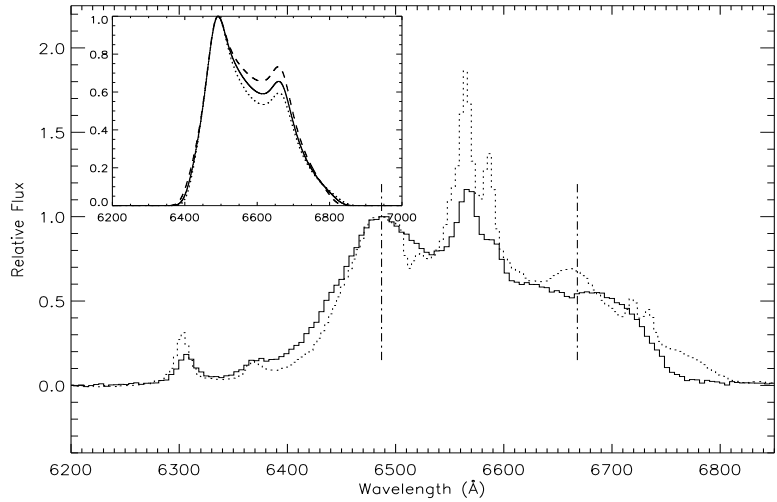}
\caption{To compare the observed line profile by HST in 2000 and mean 
observed line profile around 1995. Thin dotted line represents the mean 
spectrum around 1995 and thick solid line represents the observed line 
profile by HST in 2000. Two vertical dot-dashed lines mark the positions 
of red peak and blue peak of double-peaked broad H$\alpha$. Top left 
panel shows the theoretical results. 
Solid line represents the mean observed double-peaked line profile 
around 1995, thick dashed line represents the expected line profile 
around 2000 with clockwise rotating disk-like BLRs, thick dotted line 
represents the expected line profile around 2000 with anticlockwise 
rotating disk-like BLRs. }
\label{2000}
\end{figure*}

    We simply assume that orientation angle is the unique parameter to vary 
from 1995 to 2000 in the elliptical accretion disk model. As what we have 
done in section 3, the elliptical disk-like BLRs are separated into 60 rings. 
It is clear that orientation angle of each ring in 2000 should be some 
different due to different semi-major axis length for each ring, 
$\phi_{2000,i}=\phi_{1995,i}\pm 2\times\pi\times\frac{5}{T_{pre,i}}$, 
where $\pm$ means the rotating direction of disk-like BLRs is clockwise 
or anticlockwise for observer. After the orientation angles of the 60 rings 
around 2000 are determined, line profile expected in 2000 can be re-constructed 
by the sum of 60 expected double-peaked components from 60 rings, as what we 
have done in Section 3. Here the theoretical line profile in 2000 is mainly 
based on the mean spectrum around 1995 in AGN WATCH project. The mean 
observed spectrum shown as thin solid line in Figure~\ref{2000} is created by 
PCA (Principal Components Analysis, so-called Karhunen-Loeve Transform) method 
applied for all the 66 observed spectra of broad H$\alpha$ around 1995. The 
convenient and public IDL PCA program 'pca\_solve.pro' written by D. Schlegel 
in Princeton University is used, which is included in the SDSS software 
package of IDLSPEC2D (http://spectro.princeton.edu/). Commonly, the first 
principal component represents the mean spectrum. From the mean spectrum, it 
is clear that there are similar flux ratio ($\sim1.46$) of blue peak to red 
peak of broad double-peaked H$\alpha$ and similar flux ratio ($\sim1$) of 
blue part to red part with those mean values shown in Figure~\ref{var}.

    Figure \ref{2000} shows the probable observed line profile of 
broad double-peaked H$\alpha$ of 3C390.3 around 2000, and shows the comparison 
between the mean observed spectrum around 1995 and observed spectrum around 
2000. Based on the theoretical accretion disk model, through the results 
shown in top-left panel in Figure~\ref{2000}, we can find that due to 
precession period, after about 5 years, the flux ratio of red peak to blue 
peak should be changed. In the figure, flux density at blue peaks of all the 
line profiles have been normalized to 1. If rotating direction of disk-like BLRs 
is clockwise to observer, the expected flux ratio (1.36) of red peak to 
blue peak of spectrum around 2000 should be smaller than the one (1.46) of 
observed spectrum around 1995. If rotating direction of disk-like BLRs is 
anticlockwise to observer, the expected flux ratio (1.68) in 2000 should 
be larger than the one (1.46) in 1995. Through the observed spectrum in 
2000, the flux ratio of blue peak to red peak is about $1.79\pm0.18$. Thus, 
if anticlockwise rotating disk-like BLRs are applied for 3C390.3, the 
variations in observed spectra in different years can be naturally explained, 
although the applied model is over-simplified. The some larger value 1.79 
than theoretical expected 1.68 are perhaps due to the not so accurate 
precession period, the probable existence of varying of hot spots and/or 
warped structures, different local broadening velocity (which should lead the 
part around red peak to be some flat), etc.. The not so consistent red peak 
positions for spectrum around 2000 and mean spectrum around 1995 is probable 
due to the variation of emissivity power slope during the 5 years from 1995 
to 2000. 

     Finally, a simple summary is listed as follows. Based on the observed 
spectra around 1995 for 3C390.3, the elliptical accretion disk model is 
applied to fit the observed double-peaked broad H$\alpha$ collected from 
AGN WATCH project. Then based on the disk parameters, the formed geometrical 
structures of BLRs of 3C390.3 are applied to check the reverberation 
mapping results for 3C390.3. The disk-like BLRs supposed by theoretical 
elliptical accretion disk model can well re-produce the observational 
results about reverberation mapping technique. Thus the disk-like BLRs are 
preferred for 3C390.3. Furthermore, we check the effects of precession 
of elliptical disk-like BLRs on the observed line profile. After about 
5 years, the line profile of broad double-peaked H$\alpha$ observed 
in 2000 by HST can be well expected by the theoretical elliptical accretion 
disk model, based on the disk parameters measured through spectra 
observed around 1995, which confirms the elliptical accretion disk 
model is appropriate to double-peaked emitter 3C390.3	. 

\section*{Acknowledgements}
ZXG gratefully acknowledge the anonymous referee for 
giving us constructive comments and suggestions to greatly 
improve our paper. ZXG gratefully acknowledges finance support 
from Chinese grant NSFC-11003034, and the useful discussions from 
Prof. Ting-Gui Wang in USTC. We thank the project of AGN WATCH 
(http://www.astronomy.ohio-state.edu/\~{}agnwatch/) to make us 
conveniently collect spectra of 3C390.3. We thank Prof. Andre 
Martel in Space Telescope Science Institute who provide public 
HST spectra of 3C390.3. We thank for the The Multimission Archive 
at STScI which is a NASA funded project to support and provide to 
the astronomical community a variety of astronomical data archives, 
with the primary focus on scientifically related data sets in the 
optical, ultraviolet, and near-infrared parts of the spectrum. 
We thank for the public SDSS IDL code provided by Dr. David Schlegel 
in Princeton University, and the public Markwardt IDL Library.  
This research has made use of the NASA/IPAC Extragalactic Database 
(NED) which is operated by the Jet Propulsion Laboratory, California 
Institute of Technology, under contract with the National Aeronautics 
and Space Administration.

\end{document}